\documentclass[aip,pop,preprint]{revtex4-1}
\usepackage{graphicx}
\usepackage{dcolumn}
\usepackage{bm}
\usepackage{float}
\usepackage{colortbl}

\begin{document}

\title{Rapid Embedded Wire Heating via Resistive Guiding of Laser-Generated Fast Electrons as a Hydrodynamic Driver}
\author{A.P.L.Robinson}
\email{alex.robinson@stfc.ac.uk}
\affiliation{Central Laser Facility, STFC Rutherford-Appleton Laboratory,Didcot, OX11 0QX, United Kingdom}
\author{H.Schmitz}
\affiliation{Central Laser Facility, STFC Rutherford-Appleton Laboratory,Didcot, OX11 0QX, United Kingdom}
\author{J.Pasley}
\affiliation{York Plasma Institute, University of York, York, YO10 5DD, United Kingdom}
\affiliation{Central Laser Facility, STFC Rutherford-Appleton Laboratory,Didcot, OX11 0QX, United Kingdom}
\begin{abstract}
Resistively guiding laser-generated fast electron beams in targets consisting of a resistive wire embedded in lower $Z$ material should allow one to rapidly heat the wire to over 100eV over a substantial distance without
strongly heating the surrounding material.  On the multi-ps timescale this can drive hydrodynamic motion in the surrounding material.  Thus ultra-intense laser solid interactions have the potential as a controlled driver
of radiation hydrodynamics in solid density material.  In this paper we assess the laser and target parameters needed to achieve such rapid and controlled heating of the embedded wire.
\end{abstract}
\maketitle

\section{Introduction}
In this paper we present numerical simulations of laser-generated fast electron transport in solids \cite{jrd1,honrubia1,lancaster1,green1} where the fast electron flow is guided by means of resistive-guiding \cite{robinson1,robinson2,kar1,ramak1,pmckenna1,schmitz1}.  The aim being to strongly heat the guiding structure (a wire) so that, on the multi-ps timescale, it will drive violent hydrodynamic motion in the surrounding material.  This involves constructing the solid target from materials of different $Z$ so that the target has in-built resistivity gradients.  Strong magnetic field growth at these gradients when there is fast electron flow will produce magnetic fields that confine the fast electrons to the higher resistivity regions.  Thus the fast electrons can be guided along paths defined by target construction.  

As the fast electrons are mostly confined to the guiding structure, the strongest target heating (via Ohmic heating and background drag on the fast electrons) will also be confined to the guiding structure.  This means that the fast electrons can rapidly (few ~ps at most) heat a well defined structure to high temperature before significant hydrodynamic motion occurs.  On the multi-ps timescale there will be both a strong hydrodynamic expansion of the structure and the radiative transport of energy into the surrouding material.  Depending on the choice of various parameters this may potentially lead to the generation of strong shocks.

The use of resistive guiding to drive radiation hydrodynamics through the controlled rapid heating of solid density material with fast electrons is interesting because there is considerable interest in high energy density hydrodynamical phenomena because of their astrophysical relevance and the need for laboratory experiments that can provide some test of astrophysical simulation codes. 
Previously, this type of rapid, controllable heating which can launch very strong shocks has been achieved using cluster media \cite{symes1}, and this has made studies of strong shocks and blast waves possible.  As the cluster media are controllable, one can carry out a wide range of different experiements in which shocks are launched in different fashions.  It is interesting to see to what extent something similar can be achieved using guided fast electron heating in solid density matter, where the densities and pressures will be considerably greater than in cluster media.  In addition to laser-irradiated cluster media we also note the work that has been done on driving shock waves in the pre-plasma of solid targets irradiated by intense laser pulses with duration less than 100~fs \cite{mondal1}.

Here we will examine the case of guiding into a simple cylinder or wire of comparable size to the laser focal spot.  The focus in this paper is solely on the heating aspect of the problem, as this is the essential requirement for shock generation.  This situation has previously been considered in terms of guiding and collimating the fast electrons, however the heating of the wire itself was not considered in much detail, nor was a more realistic fast electron divergence used in these previous studies.  Some attention has been given to target heating for driving hydrodynamics by Sentoku and co-workers \cite{sentoku_shocks}, although this did not use resistive guiding as such and considered relatively thin targets.  Hence the need to revisit this scenario in order to more properly assess the simple wire as a route to launching strong cylindrical shocks into dense material.  We will show that using shorter laser wavelengths is particularly beneficial to controlled heating.  Using a higher $Z$ wire material is beneficial only up to a certain point because of limitations imposed by low temperature resistivity effects.  In principle, the heating should scale with the product of the laser intensity and pulse duration to the power of 0.4, however the additional issue of confining the fast electrons to the wire means that longer pulse durations and commensurately lower intensities may be preferable.  Therefore this type of controlled heating prefers a set of laser parameters that is substantially different from those used in most current experiments.

\section{Theory}
It is well known that rapidly heating a well defined region of material to produce a 'hot region' with a sharp transition to the surrounding, cooler material will lead to the subsequent explosion of the 'hot region' and launch a shock into the surrounding material.  If radiative energy transport is strong then the hot region will instead become a driver of radiation hydrodynamics in the surrounding material.  In either case, one needs controlled, rapid heating of a specified region of material.  Resistive guiding of laser-generated fast electrons should be able to provide this in solid density material, as the heating is rapid ($\approx$1~ps) and well-defined by virtue of confining the fast electrons to the guide element.  The key problem is whether or not good heating can be obtained along a significant length of the wire, and it is this problem that we intend to address.

Let us start by determining the important physical parameters in the wire heating problem.  We will assume that we want to obtain both a high background electron temperature in the wire, and a high $T_el$ product as we also want to obtain as long as region of heated wire as we can.  At solid density, fast electrons with MeV energies will typically have ranges of a few millimetres.  This is just the range of the fast electrons due to the collisional drag from the cold background electrons.  In the case of a guide wire that is a few hundred microns in length, collisional stopping will not therefore be the dominant heating mechanism (although it not negligible either).  The primary mechanism that has to be considered is Ohmic heating.  Above a few hundred eV, most low to moderate $Z$ materials will exhibit Spitzer-like resistivity.  In the case of a plasma with the Spitzer resistivity, which is heated due to the passage of a fast electron beam with constant current density (thus the return current has constant current density), one can analytically integrate to obtain,

\begin{equation}
T(t) = \left[Bt_h + T_0^{5/2}\right]^{2/5},
\end{equation}
where $T$ (the background electron temperature) is in eV, $T_0$ is the initial background electron temperature in eV, and $B = 5Z\log\Lambda{j_f^2}/3en_e$.  If $T \gg T_0$ then, together with $P_w = Zn_ieT$, we obtain the following dependencies for the electron temperature and electron pressure produced in the wire:
\begin{equation}
T \propto \frac{Z^{2/5}j_f^{4/5}t_h^{2/5}}{n_e^{2/5}},
\end{equation}
and,
\begin{equation}
P_w \propto Zn_i^{3/5}j_f^{4/5}t_h^{2/5}.
\end{equation}
If one now makes the further assumptions that the fast electron current density is related to the laser via a simple energy balance,
\begin{equation}
e\beta{I_L} = j_f\bar{\epsilon_f}, 
\end{equation}
and that the fast electron temperature exhibits a 'ponderomotive' scaling with the laser parameters (i.e $\bar{\epsilon_f} = A\sqrt{I_L\lambda_L})$), then
we obtain the following dependencies for the electron pressure in the wire,
\begin{equation}
\label{fin_dep}
P_w \propto \frac{Zn_i^{3/5}\beta^{4/5}I_L^{2/5}t_h^{2/5}}{\lambda_L^{4/5}}.
\end{equation}
Note that $\beta$ is the laser to fast electron conversion efficiency.  Equation \ref{fin_dep} shows that the most significant parameters are the Z of the guiding material, the laser to fast electron conversion efficiency, and the laser wavelength.  The electron pressure and temperature that is generated only scales weakly with the laser intensity and heating time (which we shall assume is approximately equal to the laser pulse duration, i.e. $t_h = \tau_L$).  Optimizing the laser to fast electron conversion efficiency is difficult as this depends on a thorough understanding of fast electron generation which is currently incomplete.  Using different materials for the guide is a matter of target fabrication, which is relatively straightforward in comparison.  One should be cautious on this point, as the above analysis assumes the Spitzer resistivity which only applies above 100-200eV in many solids, so the scaling with Z needs some further consideration.  The scaling with laser wavelength is also quite strong, and shorter wavelengths can be produced through the use of non-linear crystals.  This also entails loss of laser energy, and thus intensity.  However due to the stronger scaling with wavelength it is likely, on the basis of equation \ref{fin_dep}, that shortening the wavelength will still be beneficial.
Equation \ref{fin_dep} also shows that intensity and heating time scale with the same power, so one should obtain the same heating for the same product of $I_L\tau_L$. 

This analysis does assume, however, that the fast electron beam is perfectly confined and guided by the guide wire.  In reality this is not possible, and some consideration must therefore be given to the quality of confinement in the guide wire.  If the characteristic fast electron half-angle is $\theta_{1/2}$ then the product of the magnitude of the azimuthal magnetic flux density ($B_\phi$) and width ($R_\phi$) of the azimuthal magnetic field required for confinement can be determined if one approximates the region of confining field to be uniform.  In this case the electron will travel on a circular trajectory of radius $R_g$, the Larmor radius, and the case of limiting confinement is the case where the circular segment just touches the far side of the confining region.  The height of the circular segment is then $R_g\left(1-\cos\theta_{1/2}\right)$, so it follows that the confinement condition is,

\begin{equation}
\label{confeqn}
B_\phi{R_\phi} > \frac{\gamma_fv_fm_e}{e}\left(1 - \cos\theta_{1/2}\right).
\end{equation}
This result was previously presented by Robinson and co-workers in \cite{robinson3}, note that here we have given a fuller derivation of this equation.  Typically $B_\phi{R_\phi}$ is about 10$^{-3}$Tm and this scales slowly with $I_L$,$\lambda_L$, $\beta$, and $\tau_L$.  If $I_L >$ 10$^{19}$Wcm$^{-2}$ and $\lambda_L =$1$\mu$m then equation \ref{confeqn} will be marginally satisfied.  At shorter $\lambda_L$ and lower $I_L$ the confinement condition is much more strongly met.  This indicates that the heating may have an even stronger dependence than equation \ref{fin_dep} indicates and that higher intensity is less favourable to strong heating than equation \ref{fin_dep} suggests.

 In the subsequent section the insights obtained from the analysis will be tested by numerical simulation.  Equation \ref{fin_dep} will be used as a guide, so we will test the dependence on $I_L\tau_L$, Z, and $\lambda_L$.  The fast electron divergence angle and wire radius will also be examined by numerical simulation.

\section{Simulation}

\subsection{Set Up}
Simulations were performed using the 3D particle hybrid code {\sc zephyros}.  The 'standard' run used was set up as follows :  A 300$\times$200$\times$200 grid was used with a 1$\mu$m cell size in the $x$-direction and a 0.5$\mu$m cell size in the $y$- and $z$-directions.  The target consisted of a CH$_2$ substrate within which a 10$\mu$m Al wire was embedded.  The wire is colinear with the $x$-axis and is centred on $y$=$z$=50$\mu$m.  The background temperature is initially set to 1eV everywhere.  The background resistivity was described by the model which closely follows Lee and More, but with the minimum electron mean free path taken to be 8$r_s$, where $r_s$ is the interatomic spacing.  The background fluid equation of state was based on the Thomas-Fermi model (this includes the ionization state).  The temporal profile of the injected fast electron beam is a top-hat function of $\tau_L =$1~ps duration, and the transverse profile is $\propto \exp\left[-(r/r_{spot})^2\right]$ with $r_{spot} =$5$\mu$m.  The injected fast electron beam models irradiation at an intensity of $I_L =$5$\times$10$^{19}$Wcm$^{-2}$, with the assumption of 30\% conversion efficiency.  The fast electron angular distribution is uniform over a cone subtended by a half angle of 50$^\circ$.  The fast electron temperature used was set to, which was chosen to model irradiation at $\lambda_L =$1$\mu$m according to the Ponderomotive Scaling proposed by Wilks,
\begin{equation}
\label{wilks}
T_f  = \mbox{0.511}\left[\sqrt{1 + \frac{I_L\lambda_L^2}{\mbox{1.38}\times\mbox{10}^{18}\mbox{Wcm}^{-2}}}-1 \right]\mbox{MeV}.
\end{equation}

In the reporting of the results we will assume this 'standard' run to be the set of parameters used, and we only state the values of parameters where they differ from this standard set-up, with the exception of run B (which is a `standard run').

\subsection{Results}
In order to quantitatively describe 'heating with depth', the integral,
\begin{equation}
I_{Tl} = \int{Tdx},
\end{equation}
was calculated for each simulation at a given time along $y=z=$50$\mu$m (i.e. the target axis).  The results are tabulated in table below.

\subsubsection{Wavelength Scaling}
A set of four simulations were carried out to test the wavelength scaling (runs A--D).  These used the standard configuration, but the wavelengths chosen were $\lambda_L$=2,1, 0.5, and 0.333$\mu$m and the fast electron temperature was set according to equation \ref{wilks}.  In figure \ref{fig:figure1} we show temperature plots in the $y$-$z$ midplane at 1.2~ps in the case of 
runs B and C ($\lambda_L =$ 1 and 0.5 $\mu$m), and in figure \ref{fig:figure2} we plot the background electron temperature at $x =$ 50,100, 150, and 200 $\mu$m ($y = z =$50$\mu$m) at 1.2~ps in all four simulations.

It is clear from figures \ref{fig:figure1} and \ref{fig:figure2} that reducing laser wavelength has a very strong effect on the heating of the wire.  The tabulated values of $I_{Tl}$ (table \ref{templ_tbl}) confirm this trend, and show that it is stronger than $\propto \lambda_L^{-4/5}$.  As the wavelength is reduced, both the absolute electron temperature and the heating with depth increase substantially.  The scaling is usually stronger than the $\propto \lambda_L^{-4/5|}$ predicted by equation \ref{fin_dep}, which indicates that the increase in the confinement of the electrons in the guide wire on decreasing $\lambda_L$ is the most important effect throughout the parameter space explored here.  The previous consideration of the confinement condition, i.e. equation \ref{confeqn} lead to this conclusion.  On inspecting the evolution of the fast electron density we find that there is visibly better confinement of the fast electrons along the length of the guide wire, as is shown in figure \ref{fig:figure3} where we show the fast electron density in the midplane at 1~ps in runs B and C.

Our numerical study of the role of the laser wavelength therefore leads us to conclude that reducing the laser wavelength below 1$\mu$m (e.g. by frequency doubling or tripling) greatly enhances the heating of the guide wire.  The enhancement is stronger than one expects in the case of ideal transport along the wire (equation \ref{fin_dep}), because shorter wavelengths also greatly enhance the confinement of fast electrons in the wire. 

\subsubsection{$I_L\tau_L$ Scaling}
A set of four simulations (B,E--G) were carried out to study the scaling with $I_L\tau_L$, which in the case of ideal transport along the wire should lead to a scaling of $(I_L\tau_L)^{2/5}$ (equation \ref{fin_dep}).  In these four simulations, the parameters were chosen to keep $I_L\tau_L$ constant, and we looked to see if the wire heating would remain approximately constant.
The pulse duration, $\tau_L$, was varied from 0.5~ps to 3~ps, and $I_L$ was varied from 10$^{20}$ to 1.66$\times$10$^{19}$Wcm$^{-2}$.  The laser wavelength used was 1$\mu$m in all these simulations.  Results of these simulations are shown in \ref{fig:figure4}, where the background electron temperature is plotted along the target axis ($y = z =$50$\mu$m) in the 25--200$\mu$m range only.  In each case, the background electron temperature is being plotted at the point in the simulation where $t = \tau_L$, i.e. just at the end of the laser irradiation.

Figure \ref{fig:figure4} shows that approximately the same heating with constant $I_L\tau_L$ is only achieved in the case of runs E and F, and even in these case there is only close agreement in certain regions.  Runs A and G show much poorer heating of the wire.  This is also reflected by the tabulated values of $I_{Tl}$ shown in table \ref{templ_tbl}.  This suggests that, as we lower the intensity and increase the pulse duration on going from run E to F, there is not a significant increase in the fast electron confinement that is achieved and the ideal transport scaling (equation \ref{fin_dep}) therefore holds reasonalbly well.  On the other hand, this also indicates that poor confinement is achieved in the case of higher intensity and shorter pulses (as noted previously), and as it expected on the basis of equation \ref{confeqn}.  

In conclusion, what we find is that, when good confinement of the fast electrons is achieved, the wire heating does not improve with constant $I_L\tau_L$ as per the ideal transport scaling (\ref{fin_dep}).  However when confinement is poor, better wire heating is achieved by reducing the laser intensity and increasing the pulse duration as this improves confinement in the wire.

\subsubsection{Effect of Fast Electron Divergence Angle}
In the standard simulation, the fast electron divergence was set to $\theta_{div} =$50$^\circ$.  This value is considerably higher than that used in a number of previous simulations.  The extent to which this affects the current results was assessed by repeating simulations B and C (simulations H and I) with $\theta_{div} =$30$^\circ$.  The resulting heating induced in these simulations at 1.2~ps is shown in figure \ref{fig:figure5}, and this figure can be directly compared to figure \ref{fig:figure1}.  

It is clear that the lower divergence fast electron beam used in these simulations has led to substantially better heating (and better heating with depth in particular).  This is also reflected by the tabulated values of $I_{Tl}$ shown in table \ref{templ_tbl}.  In this particular configuration, it is significantly easier to strongly confine the fast electrons with lower divergence angle, and this is clearly shown in equation \ref{confeqn}.  This is the principal cause of the the substantial improvement in heating with depth, as it leads to good confinement being achieved over a much greater length of the wire.

The fast electron divergence is one parameter that does not currently appear to be controllable experimentally, or at least not through manipulation of the optical parameters or basic aspects of the target.  More advanced exploitation of resistive guiding may help improve matters, and this will be subject of future work.

\subsubsection{Effect of Wire Radius}
In the standard simulation the wire radius was set to 10$\mu$m which is twice the characteristic spot radius.  Two further simulations (J and K) were run to assess the effect of using a wire with the radius matched to the injection spot radius.  These are identical to runs B and C, but with the wire radius set to 5$\mu$m instead of 10$\mu$m.  The background electron temperature is shown in figure \ref{fig:figure6} at 1.2~ps.  This figure can be directly compared to figure \ref{fig:figure1}.  

From figure \ref{fig:figure6} (and table \ref{templ_tbl}), it is clear that much better heating with depth is obtained in the case of the thinner wire in both simulations.  The reason for this is that the narrower pipe maintains a higher fast electron current density which helps both confinement and heating.  As the fast electrons will generally fill the pipe uniformly, wider pipes result in lower fast electron current density which reduces the rate at which confining magnetic field is generated at the head of the beam, as well as reducing the heating rate. 

\subsubsection{Effect of Target $Z$ and Low Temperature Resistivity}
Further simulations were carried out to analyze the effect of the linear scaling of heating (in the ideal transport case) with $Z$ (runs B,L--N, and C,O--Q).  Including run B/C, this means that C ($Z =$6), Al ($Z =$13), Ti ($Z =$22), and Cu ($Z =$29) were all tested at both $\lambda_L =$1~$\mu$m and $\lambda_L =$0.5~$\mu$m .  The background electron temperature at 1.5~ps is shown for all four cases at $\lambda_L =$1~$\mu$m  in figure \ref{fig:figure7} and for all four cases at $\lambda_L =$0.5~$\mu$m in figure \ref{fig:figure8}.  In the case of \ref{fig:figure7} we have plotted the temperature logarithmically to make the differences between the different simulations clearer.

As can be seen from both figure \ref{fig:figure7} and figure \ref{fig:figure8}, at low $Z$, increments in $Z$ lead to clear improvements in both confinement and heating, e.g. improvement from C to Al.  However at higher $Z$ one finds that the heating of the wire does not improve rapidly (e.g. Al to Ti), and the heating can then get substantially worse (e.g. Cu).  The tabulated values of $I_{Tl}$ shown in table \ref{templ_tbl} also show this trend.  The behaviour at higher $Z$ appears curious at first, but the $Z$-scaling has been derived on the basis of the Spitzer resistivity where $\eta \propto Z$.  As most of the target are only heated to a few hundred eV, with only the region which is quite close the injection region being heated above 500~eV, the heating and fast electron dynamics far from the injection region will be strongly dependent on the low temperature resistivity.  The resistivity curves used for Ti and Cu peak at higher temperature than the resistivity curve for CH, which means that there is a substantial temperature range over which the guide and substrate resisivities are inverted.  By this we mean that the resistivity of the guide wire is lower than the resistivity of the surrounding substrate material in this temperature range.  In the case of resistivity inversion, magnetic fields will grow which act to expel fast electrons from the guide wire.  Consequently this leads to poor confinement if any, and the inability to effectively heat the wire.  Clearly this result does depend on the resistivity model used, and future work will have to closely examine the accuracy of resistivity curves.  Nonetheless this does indicate the benefits that might be derived from using higher $Z$ guide elements may well become limited beyond a certain point, and this appears to be true even when the laser wavelength is reduced.  

The role of low temperature resistivity was tested by carrying out further runs (R--T) in which the minimum mean free path of the background electrons was set to 2$r_s$ instead of 8$r_s$.  The temperature profiles at 1.5~ps are shown in figure \ref{fig:figure9}, which should be compared with figure \ref{fig:figure7}.  Clearly much more suitable heating profiles have been obtained in this case, and this is also reflected by the larger values of $I_{Tl}$ that are obtained in runs R--T than in runs L--N (see table \ref{templ_tbl}).  The reduction of the minimum mean free path of the background electrons will act to increase the peak resistivity and shift the peak to lower temperatures.  This reduces the temperature range over which any inversion of resistivities occurs, and thus leads to much better confinement.  

This demonstrates the importance of low temperature resistivity in effective wire heating.  Clearly the heating patterns produced in figure \ref{fig:figure9} are far more suitable as a hydrodynamic driver than the corresponding ones shown in figure \ref{fig:figure7}.  Can we go beyond this and realize these clearly better resistivity curve through appropriate choice of material?  The standard form of Ti and Cu at room temperature is a polycrystalline metal, and this lattice structure will mean that the appropriate minimum mean free path is a substantial multiple of the interatomic spacing.  Thus the heating profile shown in figure \ref{fig:figure7} is more realistic than \ref{fig:figure9} for standard polycrystalline Ti and Cu.  In the case of C however, there is a range of allotropes.  Good evidence for the effect of lattice structure (or not) on the resistivity curve was recently obtained by McKenna and co-workers \cite{pmckenna1} and this has been reinforced by further results obtained by MacLellan \cite{maclellan1}.  In the case of vitreous carbon, the peak resistivity is thought to reside at very low temperature due to the highly disordered arrangement of C atoms, and thus the resistivity curve used in run R ($Z =$6, C).  So choosing a disordered carbon allotrope would seem to be possible, but not for a wide range of metals of interest.  Metallic glasses do exist under standard conditions, and these are all alloys of one form or another.  Provided that the average atomic Z in the alloy is $\sim$20, such a metallic glass may be one way to optimize the wire heating.  Even so, the results shown in figure \ref{fig:figure9} suggest that very high Z wires are unlikely to produce better heating than moderate Z materials even when radiative cooling is not considered.

Although we have concluded that the higher Z wires are likely to produce poorer heating profiles than the low-moderate Z wires, this still leaves the question of whether or not it reasonable to compare C and Al in the absence of considering radiative cooling.  Achieving poor heating in the higher Z wires makes matter worse still because the electron pressure advantage that comes from raising $Z$ only occurs if the effective ionization state is close to full ionization (see equation \ref{fin_dep}).  However this will only occur in high $Z$ materials if high temperatures are achieved.  At temperatures above 400eV the moderate Z wires will be in a highly ionized state, and the wires are optically thin to a reasonable approximation.  If one considers only bremsstrahlung losses then one finds that power per unit volume which is lost as bremsstrahlung ($P_{brems} \approx 10^{18}$Wcm$^{-3}$) will be about 10--100 less than the rate of Ohmic heating in any region where the wire is being strongly heated ($P_{Ohmic} > 10^{19}$Wcm$^{-3}$).  Therefore radiative cooling should not strongly affect the heating in low-moderate Z wires.  It is thought that accounting for line radiation will not significantly change this conclusion.

\section{Conclusions}
\begin{figure}[H]
\begin{center}
\includegraphics[width = \columnwidth]{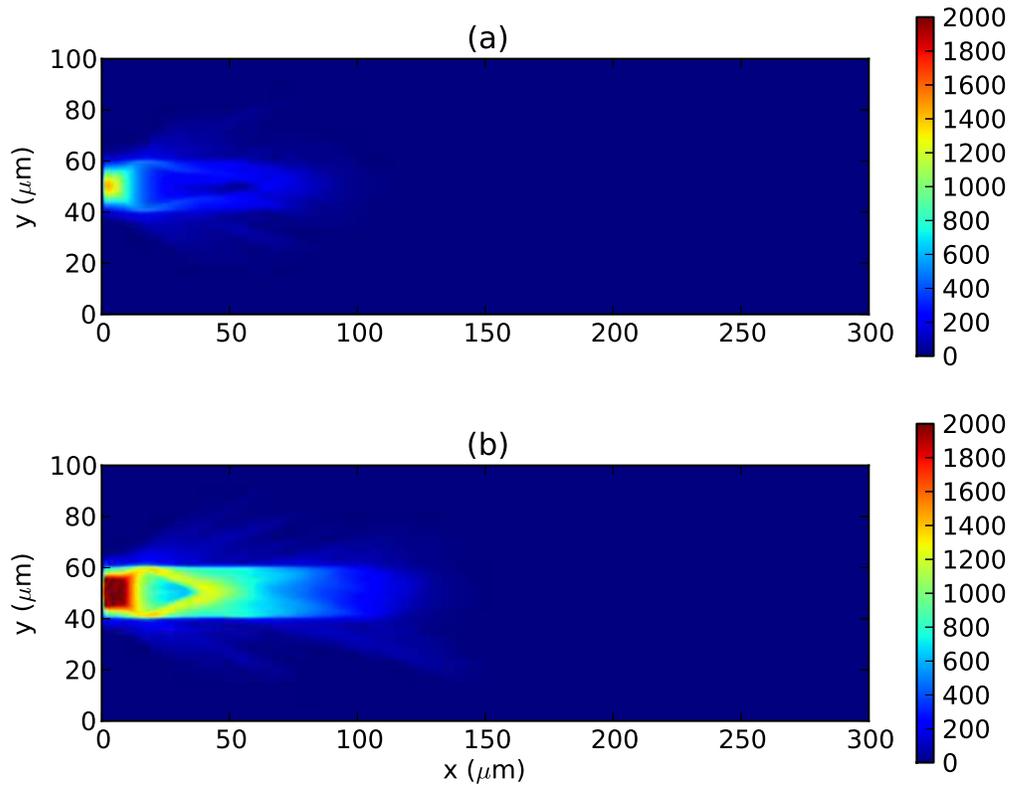} 
\caption{\label{fig:figure1} Plots of background electron temperature (eV) in $y$-$z$ midplane at 1.2~ps in (a) run B, and (b) run C. }
\end{center}
\end{figure}

\begin{figure}[H]
\begin{center}
\includegraphics[width = \columnwidth]{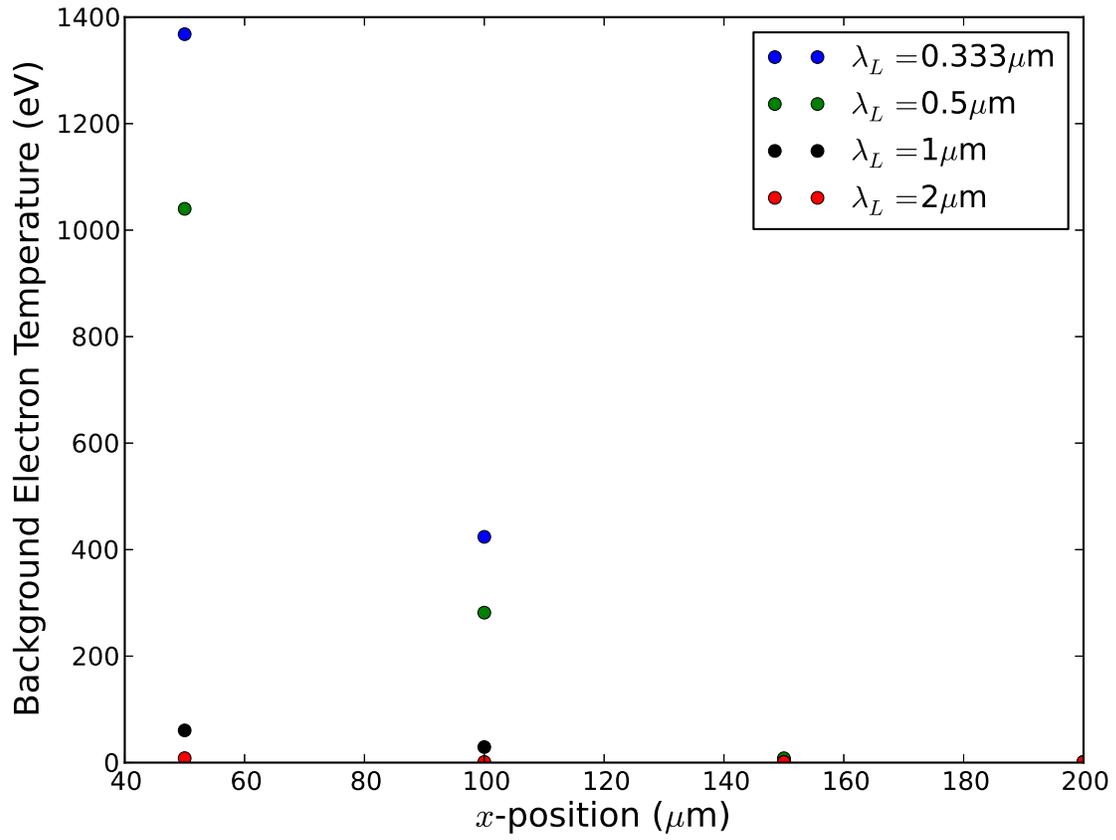} 
\caption{\label{fig:figure2} Background electron temperature at $y = z =$100$\mu$m for four $x$-positions at 1.2~ps in runs A--D. }
\end{center}
\end{figure}

\begin{figure}[H]
\begin{center}
\includegraphics[width = \columnwidth]{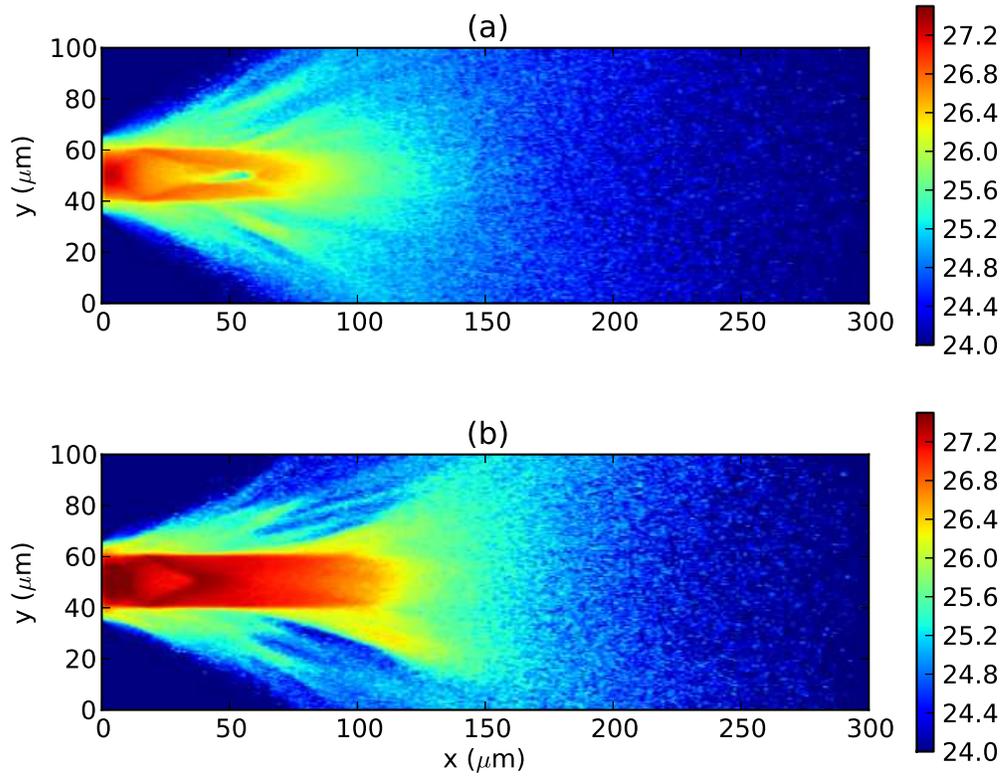} 
\caption{\label{fig:figure3} $\log_{10}$ plot of fast electron density in $y$-$z$ midplane at 1~ps in (a) run B, and (b) run C. }
\end{center}
\end{figure}

\begin{figure}[H]
\begin{center}
\includegraphics[width = \columnwidth]{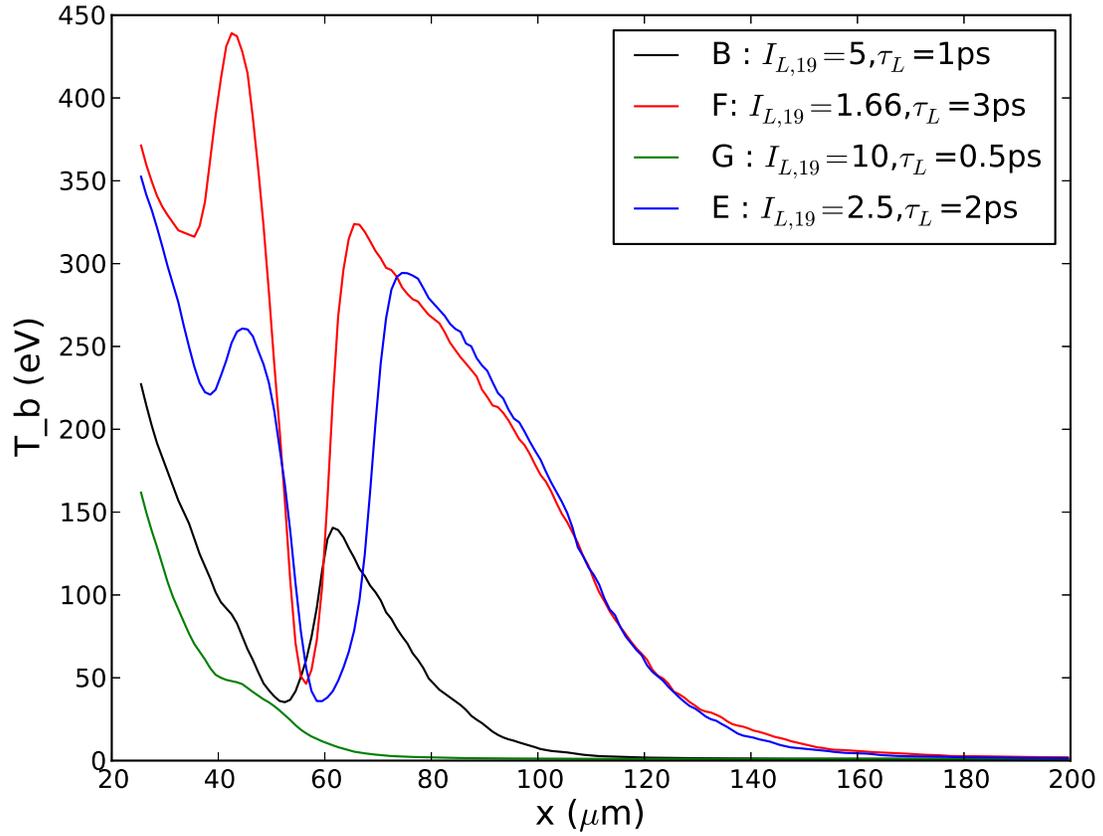} 
\caption{\label{fig:figure4} Plots of background electron temperature (eV) in runs B,E,F, and G along $y = z = $50$\mu$m for $25 \le x \le 200 \mu$m at $t=\tau_L$ in each case.}
\end{center}
\end{figure}

\begin{figure}[H]
\begin{center}
\includegraphics[width = \columnwidth]{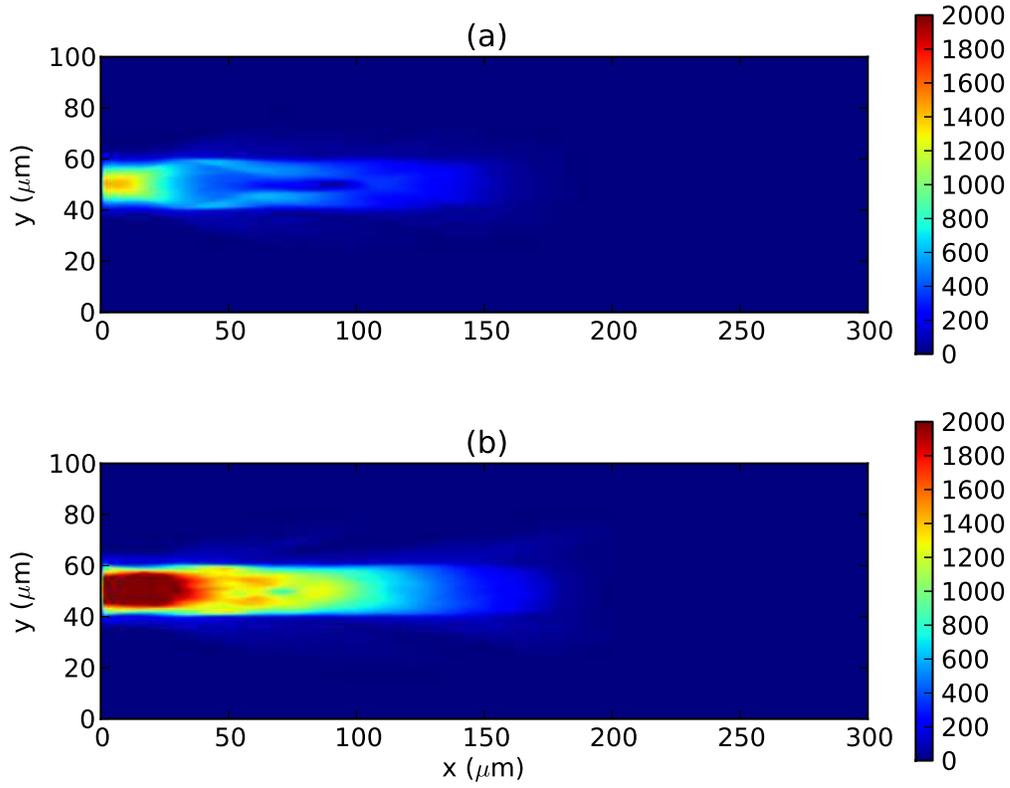} 
\caption{\label{fig:figure5} Plots of background electron temperature (eV) in $y$-$z$ midplane at 1.2~ps in (a) run H, and (b) run I.  These are repeats of runs B and C,
but with $\theta_{div} =$30$^\circ$ instead of $\theta_{div} =$50$^\circ$.}
\end{center}
\end{figure}

\begin{figure}[H]
\begin{center}
\includegraphics[width = \columnwidth]{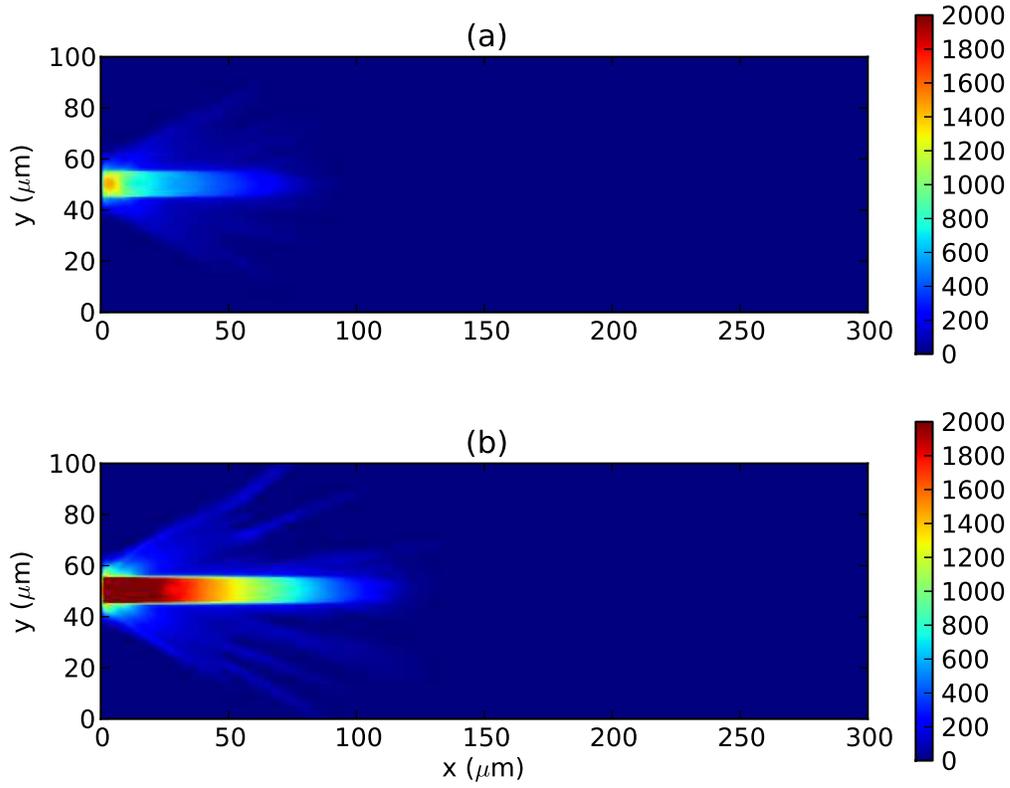} 
\caption{\label{fig:figure6} Plots of background electron temperature (eV) in $y$-$z$ midplane at 1.2~ps in (a) run J, and (b) run K.  These are repeats of runs B and C,
but with $r_{pipe} =$5$\mu$m instead of $r_{pipe} =$10$\mu$m.}
\end{center}
\end{figure}

\begin{figure}[H]
\begin{center}
\includegraphics[width = \columnwidth]{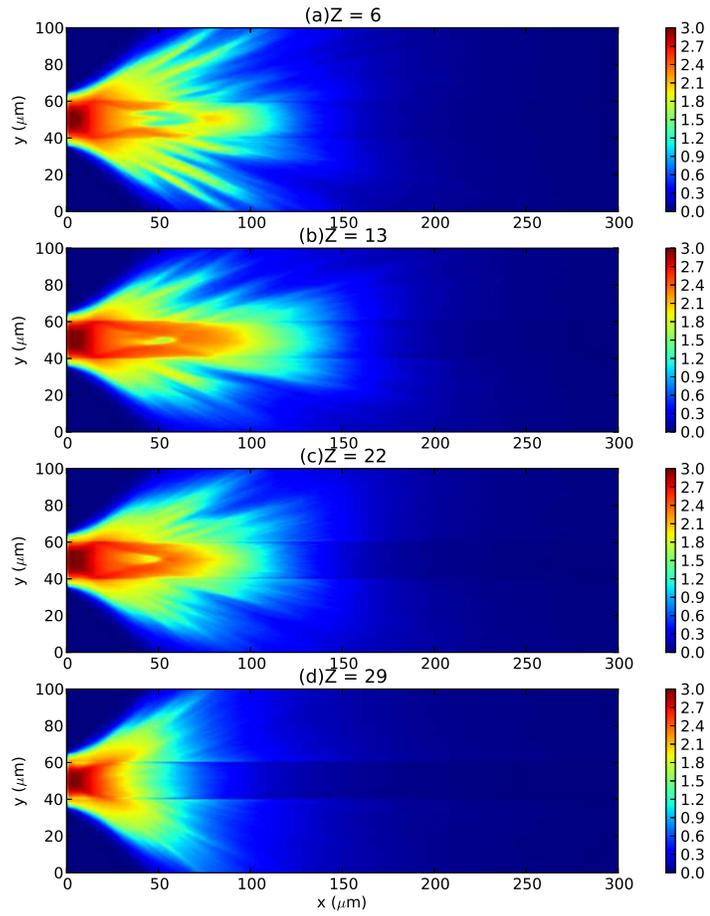} 
\caption{\label{fig:figure7} Plots of $\log_{10}$ background electron temperature (eV) in $y$-$z$ midplane at 1.5~ps in runs L(a),B(b),M (c), and N(d).}
\end{center}
\end{figure}

\begin{figure}[H]
\begin{center}
\includegraphics[width = \columnwidth]{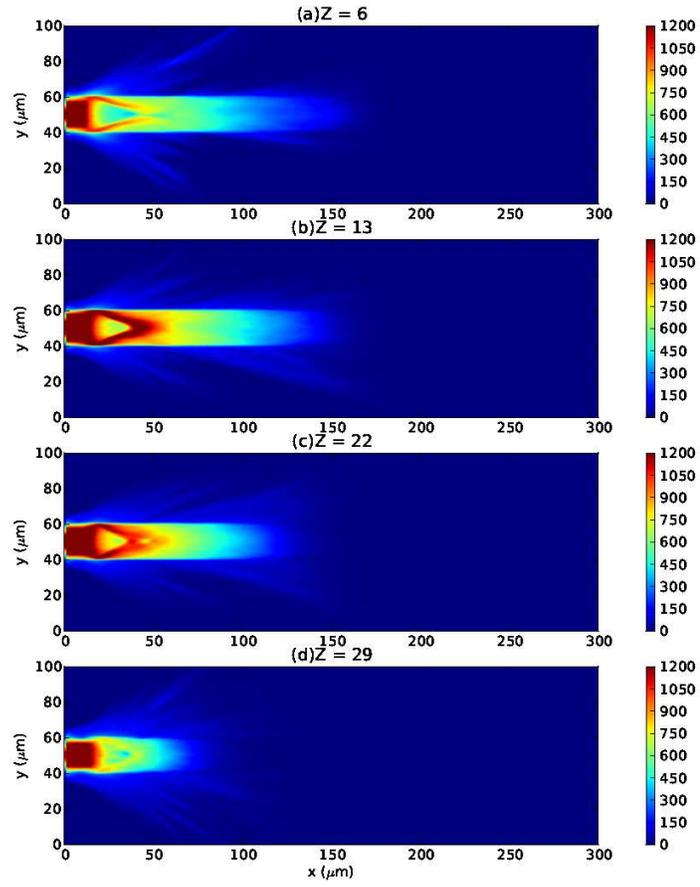} 
\caption{\label{fig:figure8} Plots of background electron temperature (eV) in $y$-$z$ midplane at 1.5~ps in runs O(a),C(b),P (c), and Q(d).  Note that in these runs $\lambda_L =$0.5~$\mu$m.}
\end{center}
\end{figure}

\begin{figure}[H]
\begin{center}
\includegraphics[width = \columnwidth]{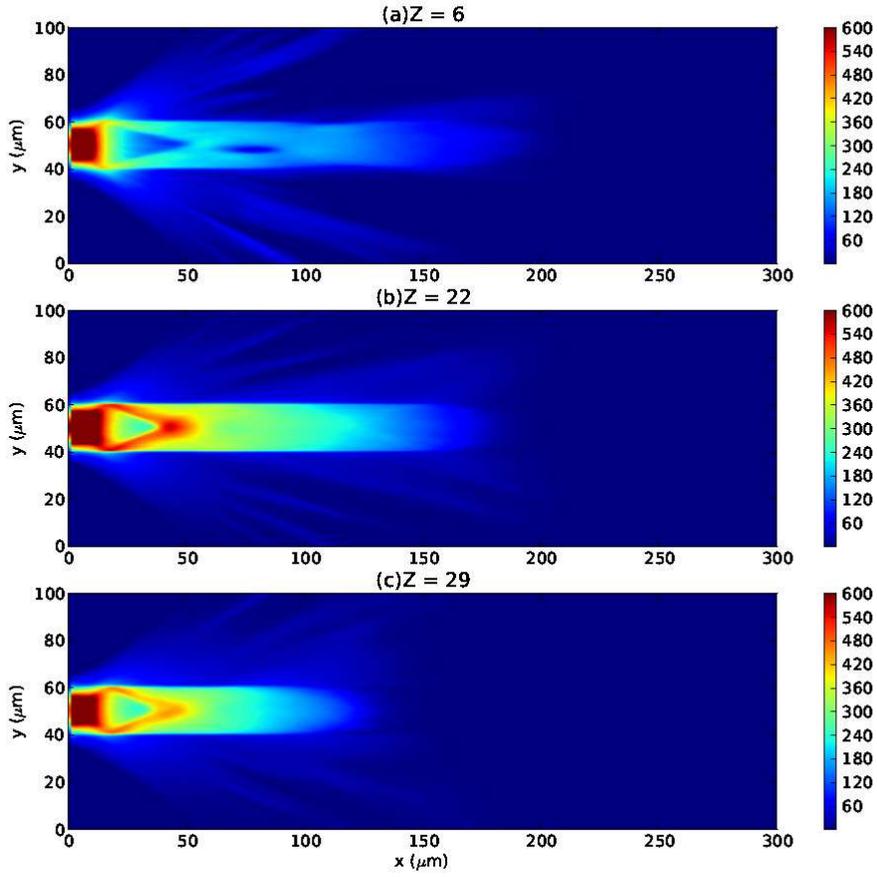} 
\caption{\label{fig:figure9} Plots of background electron temperature (eV) in $y$-$z$ midplane at 1.5~ps in runs R(a),S(b), and T(c).  Note that in these runs that the minimum electron mean free path of the background electrons was set
to 2$r_s$ instead of 8$r_s$.}
\end{center}
\end{figure}

In this paper we have studied the prospect of exploiting the resistive guiding of fast electrons to heat a 'guide element' for the purposes of driving radiation hydrodynamics on the multi-ps timescale.  In this study we have only looked at the heating problem, and we have not studied the subsequent hydrodynamic evolution.
The problem was studied from both simple analytic considerations (looking at both ideal transport along the wire and confinement), and 3D hybrid simulations  It is evident from the heating patterns that we have obtained that we can produce a well defined 'hot' region surrounded by relatively cool material with a strong gradient.  These conditions are suitable for generating a shock into the surrounding material, however there is considerable variation in the quality and extent of this hot region.  By varying different parameters we have attempted to find the parameters that lead to the greatest improvements in the heating profile.  From both of these studies we have drawn the following conclusions:
\begin{itemize}
\item{{\bf Laser Wavelength}:In the limit of ideal transport with perfect confinement, the electron pressure and temperature should scale as $\lambda_L^{-4/5}$.  The issue of confinement makes the dependence on $\lambda_L$ much stronger however, as the hybrid simulations demonstrate.  Thus going from 1$\mu$m to 0.5$\mu$m can improve the $Tl$ measure of wire heating by a factor of 3.}
\item{{\bf Intensity and Pulse Duration}:If ideal transport holds then the electron pressure and temperature should scale as $(I_L\lambda_L)^{2/5}$.  However the hybrid simulations show that
this only holds once one is in a regime where the fast electrons are well confined in the wire.  Good confinement is favoured by lower intensity (at a given wavelength) and longer pulse duration.}
\item{{\bf Fast Electron Divergence}:It is now thought that fast electron divergence angles are probably quite large ($>$45$^\circ$) in many laser-target configurations.  The heating with depth is greatly improved if the fast electron divergence angle is taken to be 30$^\circ$, as was assumed in previous simulations.  We attribute this to the fast electron beam being easier to confine at lower divergence angles.  Currently there is no clear route to reducing the intrinsic divergence angle of the fast electron beam.}
\item{{\bf Wire Radius}: Using a wire with a larger radius reduces confinement and heating as the fast electrons uniformly fill the pipe leading to a reduction in the fast electron current density.  Therefore one should not use a wire with a radius that is much greater than the laser spot radius.}
\item{{\bf Wire $Z$ and Low Temperature Resistivity}: If one assumes that the resistivity closely follows the Spitzer resistivity, then both wire heating and confinement of the fast electrons in the wire should improve linearly with the $Z$ of the wire material.  However the effect of resistivity at low temperatures (< 100~eV) may well limit the efficacy of higher $Z$ materials.  Future work will have to examine this in more detail by using successively more accurate resistivity models.  Very substantial improvements in wire heating may come from using amorphous (disordered) materials such as vitreous carbon and metallic glasses rather than polycrystalline materials. }
\end{itemize}
Overall it is clear that attempting to use laser and target parameters that are typical to current high-energy short-pulse systems (e.g. VULCAN : 1~ps, $I >$1$\times$10$^{20}$Wcm$^{-2}$,$\lambda =$1$\mu$m and polycrystalline metals) will produce relatively poor heating profiles (see figure \ref{fig:figure1}(a)).  Through frequency conversion to 2$\omega_L$ or even 3$\omega_L$ and possibly also increasing the pulse duration at the expense of intensity, it is likely that much better heating profiles can be obtained.  Much better results can also be obtained through use of amorphous materials for the wire.  If this approach to developing a radiation hydrodynamics driver in solid density material is to be pursued experimentally, then these results strongly indicate that this is direction that should be taken.

\begin{acknowledgements}
This work was supported by the European Research Council's  STRUCMAGFAST grant (ERC-StG-2012).  APLR and HS are grateful for the use of computing resources provided by STFC's Scientific Computing Department.  JP was supported by 
EPSRC grant EP/I030018/1.

\end{acknowledgements}


\begin{thebibliography}{14}%
\makeatletter
\providecommand \@ifxundefined [1]{%
 \@ifx{#1\undefined}
}%
\providecommand \@ifnum [1]{%
 \ifnum #1\expandafter \@firstoftwo
 \else \expandafter \@secondoftwo
 \fi
}%
\providecommand \@ifx [1]{%
 \ifx #1\expandafter \@firstoftwo
 \else \expandafter \@secondoftwo
 \fi
}%
\providecommand \natexlab [1]{#1}%
\providecommand \enquote  [1]{``#1''}%
\providecommand \bibnamefont  [1]{#1}%
\providecommand \bibfnamefont [1]{#1}%
\providecommand \citenamefont [1]{#1}%
\providecommand \href@noop [0]{\@secondoftwo}%
\providecommand \href [0]{\begingroup \@sanitize@url \@href}%
\providecommand \@href[1]{\@@startlink{#1}\@@href}%
\providecommand \@@href[1]{\endgroup#1\@@endlink}%
\providecommand \@sanitize@url [0]{\catcode `\\12\catcode `\$12\catcode
  `\&12\catcode `\#12\catcode `\^12\catcode `\_12\catcode `\%12\relax}%
\providecommand \@@startlink[1]{}%
\providecommand \@@endlink[0]{}%
\providecommand \url  [0]{\begingroup\@sanitize@url \@url }%
\providecommand \@url [1]{\endgroup\@href {#1}{\urlprefix }}%
\providecommand \urlprefix  [0]{URL }%
\providecommand \Eprint [0]{\href }%
\providecommand \doibase [0]{http://dx.doi.org/}%
\providecommand \selectlanguage [0]{\@gobble}%
\providecommand \bibinfo  [0]{\@secondoftwo}%
\providecommand \bibfield  [0]{\@secondoftwo}%
\providecommand \translation [1]{[#1]}%
\providecommand \BibitemOpen [0]{}%
\providecommand \bibitemStop [0]{}%
\providecommand \bibitemNoStop [0]{.\EOS\space}%
\providecommand \EOS [0]{\spacefactor3000\relax}%
\providecommand \BibitemShut  [1]{\csname bibitem#1\endcsname}%
\let\auto@bib@innerbib\@empty
\bibitem [{\citenamefont {J.R.Davies}(2002)}]{jrd1}%
  \BibitemOpen
  \bibfield  {author} {\bibinfo {author} {\bibnamefont {J.R.Davies}},\
  }\href@noop {} {\bibfield  {journal} {\bibinfo  {journal} {Phys.Rev.E}\
  }\textbf {\bibinfo {volume} {{\bf 65}}},\ \bibinfo {pages} {026407} (\bibinfo
  {year} {2002})}\BibitemShut {NoStop}%
\bibitem [{\citenamefont {J.J.Honrubia}\ and\ \citenamefont {ter
  Vehn}(2009)}]{honrubia1}%
  \BibitemOpen
  \bibfield  {author} {\bibinfo {author} {\bibnamefont {J.J.Honrubia}}\ and\
  \bibinfo {author} {\bibfnamefont {J.}~\bibnamefont {ter Vehn}},\ }\href@noop
  {} {\bibfield  {journal} {\bibinfo  {journal} {Plasma Phys.Control.Fusion}\
  }\textbf {\bibinfo {volume} {51}},\ \bibinfo {pages} {014008} (\bibinfo
  {year} {2009})}\BibitemShut {NoStop}%
\bibitem [{\citenamefont {K.L.Lancaster}\ \emph {et~al.}(2007)\citenamefont
  {K.L.Lancaster}, \citenamefont {J.S.Green}, \citenamefont {D.S.Hey},
  \citenamefont {K.U.Akli}, \citenamefont {J.R.Davies}, \citenamefont
  {R.J.Clarke}, \citenamefont {R.R.Freeman}, \citenamefont {H.Habara},
  \citenamefont {M.H.Key}, \citenamefont {R.Kodama}, \citenamefont
  {K.Krushelnick}, \citenamefont {C.D.Murphy}, \citenamefont {M.Nakatsutsumi},
  \citenamefont {P.Simpson}, \citenamefont {C.Stoeckl}, \citenamefont
  {T.Yabuuchi}, \citenamefont {M.Zepf},\ and\ \citenamefont
  {P.A.Norreys}}]{lancaster1}%
  \BibitemOpen
  \bibfield  {author} {\bibinfo {author} {\bibnamefont {K.L.Lancaster}},
  \bibinfo {author} {\bibnamefont {J.S.Green}}, \bibinfo {author} {\bibnamefont
  {D.S.Hey}}, \bibinfo {author} {\bibnamefont {K.U.Akli}}, \bibinfo {author}
  {\bibnamefont {J.R.Davies}}, \bibinfo {author} {\bibnamefont {R.J.Clarke}},
  \bibinfo {author} {\bibnamefont {R.R.Freeman}}, \bibinfo {author}
  {\bibnamefont {H.Habara}}, \bibinfo {author} {\bibnamefont {M.H.Key}},
  \bibinfo {author} {\bibnamefont {R.Kodama}}, \bibinfo {author} {\bibnamefont
  {K.Krushelnick}}, \bibinfo {author} {\bibnamefont {C.D.Murphy}}, \bibinfo
  {author} {\bibnamefont {M.Nakatsutsumi}}, \bibinfo {author} {\bibnamefont
  {P.Simpson}}, \bibinfo {author} {\bibnamefont {C.Stoeckl}}, \bibinfo {author}
  {\bibnamefont {T.Yabuuchi}}, \bibinfo {author} {\bibnamefont {M.Zepf}}, \
  and\ \bibinfo {author} {\bibnamefont {P.A.Norreys}},\ }\href@noop {}
  {\bibfield  {journal} {\bibinfo  {journal} {Phys.Rev.Lett.}\ }\textbf
  {\bibinfo {volume} {98}},\ \bibinfo {pages} {125002} (\bibinfo {year}
  {2007})}\BibitemShut {NoStop}%
\bibitem [{\citenamefont {J.S.Green}\ \emph {et~al.}(2008)\citenamefont
  {J.S.Green}, \citenamefont {V.M.Ovchinnikov}, \citenamefont {R.G.Evans},
  \citenamefont {K.U.Akli}, \citenamefont {H.Azechi}, \citenamefont {F.N.Beg},
  \citenamefont {C.Bellei}, \citenamefont {R.R.Freeman}, \citenamefont
  {H.Habara}, \citenamefont {R.Heathcote}, \citenamefont {M.H.Key},
  \citenamefont {J.A.King}, \citenamefont {K.L.Lancaster}, \citenamefont
  {N.C.Lopes}, \citenamefont {T.Ma}, \citenamefont {A.J.MacKinnon},
  \citenamefont {K.Markey}, \citenamefont {A.McPhee}, \citenamefont
  {Z.Najmudin}, \citenamefont {P.Nilson}, \citenamefont {R.Onofrei},
  \citenamefont {R.Stephens}, \citenamefont {K.Takeda}, \citenamefont
  {K.A.Tanaka}, \citenamefont {W.Theobald}, \citenamefont {T.Tanimoto},
  \citenamefont {J.Waugh}, \citenamefont {Woerkom}, \citenamefont
  {N.C.Woolsey}, \citenamefont {M.Zepf}, \citenamefont {J.R.Davies},\ and\
  \citenamefont {P.A.Norreys}}]{green1}%
  \BibitemOpen
  \bibfield  {author} {\bibinfo {author} {\bibnamefont {J.S.Green}}, \bibinfo
  {author} {\bibnamefont {V.M.Ovchinnikov}}, \bibinfo {author} {\bibnamefont
  {R.G.Evans}}, \bibinfo {author} {\bibnamefont {K.U.Akli}}, \bibinfo {author}
  {\bibnamefont {H.Azechi}}, \bibinfo {author} {\bibnamefont {F.N.Beg}},
  \bibinfo {author} {\bibnamefont {C.Bellei}}, \bibinfo {author} {\bibnamefont
  {R.R.Freeman}}, \bibinfo {author} {\bibnamefont {H.Habara}}, \bibinfo
  {author} {\bibnamefont {R.Heathcote}}, \bibinfo {author} {\bibnamefont
  {M.H.Key}}, \bibinfo {author} {\bibnamefont {J.A.King}}, \bibinfo {author}
  {\bibnamefont {K.L.Lancaster}}, \bibinfo {author} {\bibnamefont {N.C.Lopes}},
  \bibinfo {author} {\bibnamefont {T.Ma}}, \bibinfo {author} {\bibnamefont
  {A.J.MacKinnon}}, \bibinfo {author} {\bibnamefont {K.Markey}}, \bibinfo
  {author} {\bibnamefont {A.McPhee}}, \bibinfo {author} {\bibnamefont
  {Z.Najmudin}}, \bibinfo {author} {\bibnamefont {P.Nilson}}, \bibinfo {author}
  {\bibnamefont {R.Onofrei}}, \bibinfo {author} {\bibnamefont {R.Stephens}},
  \bibinfo {author} {\bibnamefont {K.Takeda}}, \bibinfo {author} {\bibnamefont
  {K.A.Tanaka}}, \bibinfo {author} {\bibnamefont {W.Theobald}}, \bibinfo
  {author} {\bibnamefont {T.Tanimoto}}, \bibinfo {author} {\bibnamefont
  {J.Waugh}}, \bibinfo {author} {\bibfnamefont {L.}~\bibnamefont {Woerkom}},
  \bibinfo {author} {\bibnamefont {N.C.Woolsey}}, \bibinfo {author}
  {\bibnamefont {M.Zepf}}, \bibinfo {author} {\bibnamefont {J.R.Davies}}, \
  and\ \bibinfo {author} {\bibnamefont {P.A.Norreys}},\ }\href@noop {}
  {\bibfield  {journal} {\bibinfo  {journal} {Phys.Rev.Lett.}\ }\textbf
  {\bibinfo {volume} {100}},\ \bibinfo {pages} {015003} (\bibinfo {year}
  {2008})}\BibitemShut {NoStop}%
\bibitem [{\citenamefont {A.P.L.Robinson}\ and\ \citenamefont
  {M.Sherlock}(2007)}]{robinson1}%
  \BibitemOpen
  \bibfield  {author} {\bibinfo {author} {\bibnamefont {A.P.L.Robinson}}\ and\
  \bibinfo {author} {\bibnamefont {M.Sherlock}},\ }\href@noop {} {\bibfield
  {journal} {\bibinfo  {journal} {Phys.Plasmas}\ }\textbf {\bibinfo {volume}
  {14}},\ \bibinfo {pages} {083105} (\bibinfo {year} {2007})}\BibitemShut
  {NoStop}%
\bibitem [{\citenamefont {A.P.L.Robinson}\ \emph {et~al.}(2012)\citenamefont
  {A.P.L.Robinson}, \citenamefont {M.H.Key},\ and\ \citenamefont
  {M.Tabak}}]{robinson2}%
  \BibitemOpen
  \bibfield  {author} {\bibinfo {author} {\bibnamefont {A.P.L.Robinson}},
  \bibinfo {author} {\bibnamefont {M.H.Key}}, \ and\ \bibinfo {author}
  {\bibnamefont {M.Tabak}},\ }\href@noop {} {\bibfield  {journal} {\bibinfo
  {journal} {Phys.Rev.Lett.}\ }\textbf {\bibinfo {volume} {108}},\ \bibinfo
  {pages} {125004} (\bibinfo {year} {2012})}\BibitemShut {NoStop}%
\bibitem [{\citenamefont {S.Kar}\ \emph {et~al.}(2009)\citenamefont {S.Kar},
  \citenamefont {A.P.L.Robinson}, \citenamefont {D.C.Carroll}, \citenamefont
  {O.Lundh}, \citenamefont {K.Markey}, \citenamefont {P.McKenna}, \citenamefont
  {P.Norreys},\ and\ \citenamefont {M.Zepf}}]{kar1}%
  \BibitemOpen
  \bibfield  {author} {\bibinfo {author} {\bibnamefont {S.Kar}}, \bibinfo
  {author} {\bibnamefont {A.P.L.Robinson}}, \bibinfo {author} {\bibnamefont
  {D.C.Carroll}}, \bibinfo {author} {\bibnamefont {O.Lundh}}, \bibinfo {author}
  {\bibnamefont {K.Markey}}, \bibinfo {author} {\bibnamefont {P.McKenna}},
  \bibinfo {author} {\bibnamefont {P.Norreys}}, \ and\ \bibinfo {author}
  {\bibnamefont {M.Zepf}},\ }\href@noop {} {\bibfield  {journal} {\bibinfo
  {journal} {Phys.Rev.Lett.}\ }\textbf {\bibinfo {volume} {102}},\ \bibinfo
  {pages} {055001} (\bibinfo {year} {2009})}\BibitemShut {NoStop}%
\bibitem [{\citenamefont {B.Ramakrishna}\ \emph {et~al.}(2010)\citenamefont
  {B.Ramakrishna}, \citenamefont {S.Kar}, \citenamefont {A.P.L.Robinson},
  \citenamefont {D.J.Adams}, \citenamefont {K.Markey}, \citenamefont
  {M.N.Quinn}, \citenamefont {X.H.Yuan}, \citenamefont {P.McKenna},
  \citenamefont {K.L.Lancaster}, \citenamefont {J.S.Green}, \citenamefont
  {R.H.H.Scott}, \citenamefont {P.A.Norreys}, \citenamefont {J.Schreiber},\
  and\ \citenamefont {M.Zepf}}]{ramak1}%
  \BibitemOpen
  \bibfield  {author} {\bibinfo {author} {\bibnamefont {B.Ramakrishna}},
  \bibinfo {author} {\bibnamefont {S.Kar}}, \bibinfo {author} {\bibnamefont
  {A.P.L.Robinson}}, \bibinfo {author} {\bibnamefont {D.J.Adams}}, \bibinfo
  {author} {\bibnamefont {K.Markey}}, \bibinfo {author} {\bibnamefont
  {M.N.Quinn}}, \bibinfo {author} {\bibnamefont {X.H.Yuan}}, \bibinfo {author}
  {\bibnamefont {P.McKenna}}, \bibinfo {author} {\bibnamefont {K.L.Lancaster}},
  \bibinfo {author} {\bibnamefont {J.S.Green}}, \bibinfo {author} {\bibnamefont
  {R.H.H.Scott}}, \bibinfo {author} {\bibnamefont {P.A.Norreys}}, \bibinfo
  {author} {\bibnamefont {J.Schreiber}}, \ and\ \bibinfo {author} {\bibnamefont
  {M.Zepf}},\ }\href@noop {} {\bibfield  {journal} {\bibinfo  {journal}
  {Phys.Rev.Lett.}\ }\textbf {\bibinfo {volume} {105}},\ \bibinfo {pages}
  {135001} (\bibinfo {year} {2010})}\BibitemShut {NoStop}%
\bibitem [{\citenamefont {P.McKenna}\ \emph {et~al.}(2011)\citenamefont
  {P.McKenna}, \citenamefont {A.P.L.Robinson}, \citenamefont {D.Neely},
  \citenamefont {M.P.Desjarlais}, \citenamefont {D.C.Carroll}, \citenamefont
  {M.N.Quinn}, \citenamefont {X.H.Yuan}, \citenamefont {C.M.Brenner},
  \citenamefont {M.Burza}, \citenamefont {M.Coury}, \citenamefont {P.Gallegos},
  \citenamefont {R.J.Gray}, \citenamefont {K.L.Lancaster}, \citenamefont
  {Y.T.Li}, \citenamefont {X.X.Lin}, \citenamefont {O.Tresca},\ and\
  \citenamefont {C.-G.Wahlstr\"{o}m}}]{pmckenna1}%
  \BibitemOpen
  \bibfield  {author} {\bibinfo {author} {\bibnamefont {P.McKenna}}, \bibinfo
  {author} {\bibnamefont {A.P.L.Robinson}}, \bibinfo {author} {\bibnamefont
  {D.Neely}}, \bibinfo {author} {\bibnamefont {M.P.Desjarlais}}, \bibinfo
  {author} {\bibnamefont {D.C.Carroll}}, \bibinfo {author} {\bibnamefont
  {M.N.Quinn}}, \bibinfo {author} {\bibnamefont {X.H.Yuan}}, \bibinfo {author}
  {\bibnamefont {C.M.Brenner}}, \bibinfo {author} {\bibnamefont {M.Burza}},
  \bibinfo {author} {\bibnamefont {M.Coury}}, \bibinfo {author} {\bibnamefont
  {P.Gallegos}}, \bibinfo {author} {\bibnamefont {R.J.Gray}}, \bibinfo {author}
  {\bibnamefont {K.L.Lancaster}}, \bibinfo {author} {\bibnamefont {Y.T.Li}},
  \bibinfo {author} {\bibnamefont {X.X.Lin}}, \bibinfo {author} {\bibnamefont
  {O.Tresca}}, \ and\ \bibinfo {author} {\bibnamefont {C.-G.Wahlstr\"{o}m}},\
  }\href@noop {} {\bibfield  {journal} {\bibinfo  {journal} {Phys.Rev.Lett.}\
  }\textbf {\bibinfo {volume} {106}},\ \bibinfo {pages} {185004} (\bibinfo
  {year} {2011})}\BibitemShut {NoStop}%
\bibitem [{\citenamefont {H.Schmitz}\ \emph {et~al.}(2012)\citenamefont
  {H.Schmitz}, \citenamefont {R.Lloyd},\ and\ \citenamefont
  {R.G.Evans}}]{schmitz1}%
  \BibitemOpen
  \bibfield  {author} {\bibinfo {author} {\bibnamefont {H.Schmitz}}, \bibinfo
  {author} {\bibnamefont {R.Lloyd}}, \ and\ \bibinfo {author} {\bibnamefont
  {R.G.Evans}},\ }\href@noop {} {\bibfield  {journal} {\bibinfo  {journal}
  {Plasma Phys.Control.Fusion}\ }\textbf {\bibinfo {volume} {54}},\ \bibinfo
  {pages} {085016} (\bibinfo {year} {2012})}\BibitemShut {NoStop}%
\bibitem [{\citenamefont {D.R.Symes}\ \emph {et~al.}(2010)\citenamefont
  {D.R.Symes}, \citenamefont {M.Hohenburger}, \citenamefont {J.Lazarus},
  \citenamefont {J.Osterhoff}, \citenamefont {A.S.Moore}, \citenamefont
  {R.R.Faustlin}, \citenamefont {A.D.Edens}, \citenamefont {H.W.Doyle},
  \citenamefont {R.E.Carley}, \citenamefont {A.Marocchino}, \citenamefont
  {J.P.Chittenden}, \citenamefont {A.C.Bernstein}, \citenamefont
  {E.T.Gumbrell}, \citenamefont {M.Dunne}, \citenamefont {R.A.Smith},\ and\
  \citenamefont {T.Ditmire}}]{symes1}%
  \BibitemOpen
  \bibfield  {author} {\bibinfo {author} {\bibnamefont {D.R.Symes}}, \bibinfo
  {author} {\bibnamefont {M.Hohenburger}}, \bibinfo {author} {\bibnamefont
  {J.Lazarus}}, \bibinfo {author} {\bibnamefont {J.Osterhoff}}, \bibinfo
  {author} {\bibnamefont {A.S.Moore}}, \bibinfo {author} {\bibnamefont
  {R.R.Faustlin}}, \bibinfo {author} {\bibnamefont {A.D.Edens}}, \bibinfo
  {author} {\bibnamefont {H.W.Doyle}}, \bibinfo {author} {\bibnamefont
  {R.E.Carley}}, \bibinfo {author} {\bibnamefont {A.Marocchino}}, \bibinfo
  {author} {\bibnamefont {J.P.Chittenden}}, \bibinfo {author} {\bibnamefont
  {A.C.Bernstein}}, \bibinfo {author} {\bibnamefont {E.T.Gumbrell}}, \bibinfo
  {author} {\bibnamefont {M.Dunne}}, \bibinfo {author} {\bibnamefont
  {R.A.Smith}}, \ and\ \bibinfo {author} {\bibnamefont {T.Ditmire}},\
  }\href@noop {} {\bibfield  {journal} {\bibinfo  {journal} {High Energy
  Density Phys.}\ }\textbf {\bibinfo {volume} {6}},\ \bibinfo {pages} {274}
  (\bibinfo {year} {2010})}\BibitemShut {NoStop}%
\bibitem [{\citenamefont {S.Mondal}\ \emph {et~al.}(2010)\citenamefont
  {S.Mondal}, \citenamefont {Amit.D.Lad}, \citenamefont {S.Ahmed},
  \citenamefont {V.Narayan}, \citenamefont {J.Pasley}, \citenamefont
  {P.P.Rajeev}, \citenamefont {A.P.L.Robinson},\ and\ \citenamefont
  {Kumar}}]{mondal1}%
  \BibitemOpen
  \bibfield  {author} {\bibinfo {author} {\bibnamefont {S.Mondal}}, \bibinfo
  {author} {\bibnamefont {Amit.D.Lad}}, \bibinfo {author} {\bibnamefont
  {S.Ahmed}}, \bibinfo {author} {\bibnamefont {V.Narayan}}, \bibinfo {author}
  {\bibnamefont {J.Pasley}}, \bibinfo {author} {\bibnamefont {P.P.Rajeev}},
  \bibinfo {author} {\bibnamefont {A.P.L.Robinson}}, \ and\ \bibinfo {author}
  {\bibfnamefont {G.}~\bibnamefont {Kumar}},\ }\href@noop {} {\bibfield
  {journal} {\bibinfo  {journal} {Phys.Rev.Lett.}\ }\textbf {\bibinfo {volume}
  {105}},\ \bibinfo {pages} {105002} (\bibinfo {year} {2010})}\BibitemShut
  {NoStop}%
\bibitem [{\citenamefont {Y.Sentoku}\ \emph {et~al.}(2007)\citenamefont
  {Y.Sentoku}, \citenamefont {A.J.Kemp}, \citenamefont {R.Presura},
  \citenamefont {M.S.Bakeman},\ and\ \citenamefont
  {T.E.Cowan}}]{sentoku_shocks}%
  \BibitemOpen
  \bibfield  {author} {\bibinfo {author} {\bibnamefont {Y.Sentoku}}, \bibinfo
  {author} {\bibnamefont {A.J.Kemp}}, \bibinfo {author} {\bibnamefont
  {R.Presura}}, \bibinfo {author} {\bibnamefont {M.S.Bakeman}}, \ and\ \bibinfo
  {author} {\bibnamefont {T.E.Cowan}},\ }\href@noop {} {\bibfield  {journal}
  {\bibinfo  {journal} {Phys.Plasmas}\ }\textbf {\bibinfo {volume} {14}},\
  \bibinfo {pages} {122701} (\bibinfo {year} {2007})}\BibitemShut {NoStop}%
\bibitem [{\citenamefont {A.P.L.Robinson}\ \emph {et~al.}(2008)\citenamefont
  {A.P.L.Robinson}, \citenamefont {M.Sherlock},\ and\ \citenamefont
  {P.A.Norreys}}]{robinson3}%
  \BibitemOpen
  \bibfield  {author} {\bibinfo {author} {\bibnamefont {A.P.L.Robinson}},
  \bibinfo {author} {\bibnamefont {M.Sherlock}}, \ and\ \bibinfo {author}
  {\bibnamefont {P.A.Norreys}},\ }\href@noop {} {\bibfield  {journal} {\bibinfo
   {journal} {Phys.Rev.Lett.}\ }\textbf {\bibinfo {volume} {100}},\ \bibinfo
  {pages} {025002} (\bibinfo {year} {2008})}\BibitemShut {NoStop}%
\bibitem [{\citenamefont {D.A.MacLellan}\ \emph {et~al.}(2013)\citenamefont
  {D.A.MacLellan}, \citenamefont {D.C.Carroll}, \citenamefont {R.J.Gray},
  \citenamefont {N.Booth}, \citenamefont {M.Burza}, \citenamefont
  {M.P.Desjarlais}, \citenamefont {F.Du}, \citenamefont {B.Gonzalez-Izquierdo},
  \citenamefont {D.Neely}, \citenamefont {H.W.Powell}, \citenamefont
  {A.P.L.Robinson}, \citenamefont {D.R.Rusby}, \citenamefont {G.G.Scott},
  \citenamefont {X.H.Yuan}, \citenamefont {C.-G.Wahlstr\"{o}m},\ and\
  \citenamefont {P.McKenna}}]{maclellan1}%
  \BibitemOpen
  \bibfield  {author} {\bibinfo {author} {\bibnamefont {D.A.MacLellan}},
  \bibinfo {author} {\bibnamefont {D.C.Carroll}}, \bibinfo {author}
  {\bibnamefont {R.J.Gray}}, \bibinfo {author} {\bibnamefont {N.Booth}},
  \bibinfo {author} {\bibnamefont {M.Burza}}, \bibinfo {author} {\bibnamefont
  {M.P.Desjarlais}}, \bibinfo {author} {\bibnamefont {F.Du}}, \bibinfo {author}
  {\bibnamefont {B.Gonzalez-Izquierdo}}, \bibinfo {author} {\bibnamefont
  {D.Neely}}, \bibinfo {author} {\bibnamefont {H.W.Powell}}, \bibinfo {author}
  {\bibnamefont {A.P.L.Robinson}}, \bibinfo {author} {\bibnamefont
  {D.R.Rusby}}, \bibinfo {author} {\bibnamefont {G.G.Scott}}, \bibinfo {author}
  {\bibnamefont {X.H.Yuan}}, \bibinfo {author} {\bibnamefont
  {C.-G.Wahlstr\"{o}m}}, \ and\ \bibinfo {author} {\bibnamefont {P.McKenna}},\
  }\href@noop {} {\bibfield  {journal} {\bibinfo  {journal} {Phys.Rev.Lett.}\
  }\textbf {\bibinfo {volume} {111}},\ \bibinfo {pages} {095001} (\bibinfo
  {year} {2013})}\BibitemShut {NoStop}%
\end{thebibliography}

\begin{table}
\resizebox{\columnwidth}{!}{
\begin{tabular}{|c|c|c|c|c|c|c|c|}
\hline
Simulation & $I_L$(Wcm$^{-2}$)& $\tau_L$ (ps) & $\lambda_L$ ($\mu$m) & $Z$ (of wire material) &$\theta_{div}$ ($^\circ$) & $r_{pipe}$ ($\mu$m) & Min.m.f.p.\\
\hline
A & & & 2     & & & &\\
\hline 
\rowcolor{cyan} B & & &      & & & &\\
\hline
C & & & 0.5   & & & & \\
\hline
D & & & 0.333 & & &&\\
\hline
E & 2.5$\times$10$^{19}$ & 2 &  & & &&\\
\hline
F & 1.66$\times$10$^{19}$ & 3 &  & & &&\\
\hline
G & 1$\times$10$^{20}$&0.5 & & & &&\\
\hline
H & & & & &30$^\circ$ &&\\
\hline
I & & & 0.5 & &30$^\circ$ &&\\
\hline
J & & & & & & 5 &\\
\hline
K & & & 0.5 & & & 5&\\
\hline
L & & & & 6 & & &\\
\hline
M & & & & 22 & & &\\
\hline
N & & & & 29 & & &\\
\hline
O & & & 0.5 & 6 & & &\\
\hline
P & & & 0.5 & 22 & & &\\
\hline
Q & & & 0.5 & 29 & & &\\
\hline
R & & & & 6 & & & 2$r_s$\\
\hline
S & & & & 22 & & &2$r_s$\\
\hline
T & & & & 29 & & &2$r_s$\\
\hline
\end{tabular}
}
\caption{\label{param_tbl}Table of simulation parameters.  Where a field is blank, the value from the standard run was used.
Coloration indicates that run B corresponds to the standard run.}
\end{table}

\begin{table}
\begin{tabular}{|c|c|c|}
\hline
Simulation & Time (ps) & $I_{Tl}$ (keV$\mu$m) \\
\hline
A & 1.2 & 9.6\\
\hline 
B & 1.2 & 28.8\\
\hline
C & 1.2 & 99.5\\
\hline
D & 1.2 & 167.1\\
\hline
E & 2.0 & 41.8\\
\hline
F & 3.0 & 46.5\\
\hline
G & 0.5 & 17.9\\
\hline
H & 1.2 & 68.8\\
\hline
I & 1.2 & 199.1\\
\hline
J & 1.2 & 43.8 \\
\hline
K & 1.2 & 146.3 \\
\hline
L & 1.5 & 21.4\\
\hline
M & 1.5 & 29.2\\
\hline
N & 1.5 & 17.0\\
\hline
O & 1.5 & 86.2\\
\hline
P & 1.5 & 103.1\\
\hline
Q & 1.5 & 61.1\\
\hline
R & 1.5 & 37.3\\
\hline
S & 1.5 & 56.8\\
\hline
T & 1.5 & 40.8\\
\hline
\end{tabular}
\caption{\label{templ_tbl}Table of $I_{Tl}$, the $Tl$ heating-with-depth metric obtained in each simulation.}
\end{table}

\end{document}